\documentclass[lettersize,journal]{IEEEtran}
\usepackage{amsmath,amsfonts}
\usepackage{array}

\usepackage[caption=false, font=footnotesize, labelformat=empty]{subfig}
\usepackage{textcomp}
\usepackage{stfloats}
\usepackage{url}
\usepackage{verbatim}
\usepackage{graphicx}
\usepackage{cite}
\makeatletter
\def\bstctlcite{\@ifnextchar[{\@bstctlcite}{\@bstctlcite[@auxout]}}
\def\@bstctlcite[#1]#2{\@bsphack
  \@for\@citeb:=#2\do{%
    \edef\@citeb{\expandafter\@firstofone\@citeb}%
    \if@filesw\immediate\write\csname #1\endcsname{\string\citation{\@citeb}}\fi}%
  \@esphack}
\makeatother

\DeclareMathOperator*{\argmin}{arg\,min}

\usepackage[ruled,vlined]{algorithm2e}

\usepackage{booktabs}
\usepackage{multirow}

\usepackage[T1]{fontenc}

\usepackage{calc} 

\usepackage[pdfstartview=XYZ,
bookmarks=true,
colorlinks=true,
linkcolor=blue,
urlcolor=blue,
citecolor=blue,
bookmarks=true,
linktocpage=true, 
hyperindex=true
]{hyperref}

\usepackage{orcidlink}

\begin{document}
\bstctlcite{IEEEexample:BSTcontrol} 

\title{Steganography Beyond Space-Time\\
with Chain of Multimodal AI}

\author{Ching-Chun Chang and Isao Echizen

\thanks{This work was supported in part by the Japan Society for the Promotion of Science (JSPS) under KAKENHI Grants (JP21H04907 and JP24H00732), and in part by the Japan Science and Technology Agency (JST) under CREST Grant (JPMJCR20D3) including AIP Challenge Program, AIP Acceleration Grant (JPMJCR24U3) and K Program Grant (JPMJKP24C2).
}
\thanks{C.-C. Chang and I. Echizen are with the Information and Society Research Division, National Institute of Informatics, Tokyo, Japan. I. Echizen is also with the Graduate School of Information Science and Technology, University of Tokyo, and the School of Multidisciplinary Sciences, Graduate University for Advanced Studies (SOKENDAI), Tokyo, Japan.
}
\thanks{Correspondence: C.-C. Chang (email: ccchang@nii.ac.jp)
}
}

\maketitle

\begin{abstract}
Steganography is the art and science of covert writing, with a broad range of applications interwoven within the realm of cybersecurity. As artificial intelligence continues to evolve, its ability to synthesise realistic content emerges as a threat in the hands of cybercriminals who seek to manipulate and misrepresent the truth. Such synthetic content introduces a non-trivial risk of overwriting the subtle changes made for the purpose of steganography. When the signals in both the spatial and temporal domains are vulnerable to unforeseen overwriting, it calls for reflection on what, if any, remains invariant. This study proposes a paradigm in steganography for audiovisual media, where messages are concealed beyond both spatial and temporal domains. A chain of multimodal artificial intelligence is developed to deconstruct audiovisual content into a cover text, embed a message within the linguistic domain, and then reconstruct the audiovisual content through synchronising both auditory and visual modalities with the resultant stego text. The message is encoded by biasing the word sampling process of a language generation model and decoded by analysing the probability distribution of word choices. The accuracy of message transmission is evaluated under both zero-bit and multi-bit capacity settings. Fidelity is assessed through both biometric and semantic similarities, capturing the identities of the recorded face and voice, as well as the core ideas conveyed through the media. Secrecy is examined through statistical comparisons between cover and stego texts. Robustness is tested across various scenarios, including audiovisual resampling, face-swapping, voice-cloning and their combinations.
\end{abstract}

\section{Introduction}

\IEEEPARstart{S}{teganography} is the study of covert writing, which has evolved from rudimentary arts, such as the use of invisible ink, into a sophisticated scientific discipline, interwoven within the field of cybersecurity~\cite{668971, 771065, Vigano:2024aa}. The applications of steganography are vast and varied, encompassing secret communication~\cite{Simmons1984, 4655281, 10.1145/1411328.1411349, 6197267, ziegler-etal-2019-neural, 10844282}, anti-counterfeit watermarking~\cite{650120, 687830, 771066, 771072, 771068}, provenance tracking~\cite{Pfitzmann:1996aa, 705568, 1188750} and forensic analysis~\cite{723401, 771070, 817228}, among others~\cite{Nagai:2018aa, Tancik:2020aa, 9736990}. In general, a steganographic system involves the process of embedding a message into a cover medium and then extracting the hidden message from the resulting stego medium. Steganography manifests in various forms, including visual, auditory and linguistic content. In other words, information can be concealed in various aspects of digital media, such as pixel intensities in imagery~\cite{1618698}, sound waves in audio~\cite{10.1007/3-540-61996-8_48}, synonym words in text~\cite{chang-clark-2014-practical} and frequency coefficients for transformed media~\cite{BARNI1998357}. The distortion introduced by steganographic systems is typically constrained to remain imperceptible to human senses. The fundamental objective of steganography is to keep the hidden information invisible and inaudible to all except the authorised parties who possess the key.

As the art and science of steganography continue to evolve, so too does the adversarial technology that seeks to manipulate digital media without authorisation. This evolution is particularly evident in the rise of artificial intelligence (AI), which has introduced transformative possibilities in revolutionising how we approach creativity, problem-solving and decision-making across various domains~\cite{Turing:1950aa, Mnih:2015aa, LeCun:2015aa, 7989324, Andrychowicz:2019aa, Udrescu:aa, Jumper:2021aa}. At the same time, however, AI has also unlocked potential threats in cyber-crime, exploiting vulnerabilities on an unprecedented scale~\cite{Barreno:2010aa, 43405, Kurakin:2017aa, 8119189, 8685687}. Generative AI, for instance, capable of synthesising realistic audio and video streams, has become a threat in the hands of those seeking to manipulate and misrepresent the truth~\cite{DBLP:journals/corr/KingmaW13, NIPS2014_5ca3e9b1, Sohl-Dickstein:aa, NEURIPS2020_4c5bcfec, 9878449, pmlr-v162-nichol22a, NEURIPS2022_ec795aea}. This type of synthetic content, often referred to as deepfakes, is highly convincing yet fraudulent, and can be exploited to deceive viewers and listeners, contributing to the spread of misinformation and the erosion of public trust~\cite{10.1145/2816795.2818056, Brock:2017aa, 10.1145/3197517.3201283, 10.1007/978-3-030-01261-8_41, NEURIPS2018_d86ea612, 10.1145/3292039, Chesney:2019aa, 10.1007/978-3-030-58517-4_42}. As this technology continues to advance, the potential for its misuse grows, raising concerns about the future of the cyber-world and highlighting the pressing need for countermeasures against these threats~\cite{Chesney:2019ab, NEURIPS2019_3e9f0fc9, 9157215, 10.1145/3425780, pmlr-v202-mitchell23a, 10238689}.

Consider a human-centric audiovisual content in which an individual's speech was recorded both aurally and visually. Suppose that the message is hidden within the pixel intensities or sound waves, serving an arbitrary purpose. Although steganographic systems can be designed to withstand distortion caused by common signal processing operations such as resampling, there remains a risk of unforeseen manipulations that go beyond the robustness premise. Generative AI technologies, such as face-swapping and voice-cloning, can potentially wipe out the hidden message, as the synthetic content has a non-trivial chance of overwriting the subtle changes made for the purpose of steganography. 


If the signals in both spatial and temporal domains are at risk of unforeseen manipulations, it prompts a reconsideration of what remains truly invariant. In quest for such an invariant domain, this study proposes a paradigm in steganography where messages are concealed beyond spatial and temporal dimensions. Motivated by multi-agent collaboration~\cite{NEURIPS2024_ee71a4b1}, we develop a chain of multimodal AI that deconstructs audiovisual content into cover text, embeds messages within the linguistic domain, and then reconstructs the audiovisual content by synchronising visual and auditory modalities with the stego text. In essence, our methodology transitions into the linguistic domain of audiovisual content, which is inherently invariant to direct spatiotemporal interference. The foundation of applied linguistic steganographic coding involves biasing the statistical distribution of word sampling~\cite{pmlr-v202-kirchenbauer23a}. Suppose a set of keywords (tokens) is shared between the sender and the receiver as the stego key. The message is encoded by biasing the word sampling process of a language generation model tasked with paraphrasing the given text, steering the model towards generating words from the shared set. The message can later be decoded by analysing the statistical likelihood of word choices, in comparison to the expected distribution of the shared set. The accuracy of message transmission is evaluated in both zero-bit and multi-bit capacity settings. In contrast to the perceptual similarity typically adopted in prior work, fidelity is extended to encompass both biometric and semantic similarities, representing identities of the recorded face and voice, as well as the ideas conveyed through the media. Secrecy is assessed through statistical analysis of cover and stego texts. Robustness is tested under conditions such as audiovisual resampling, face-swapping, voice-cloning, and combined scenarios.

The remainder of this paper is organised as follows. Section~\hyperref[sec:pre]{Preliminaries} outlines the scope of this study in relation to the primary properties of steganography. Section~\hyperref[sec:method]{Methodology} presents the proposed methodology, which involves a chain of multimodal AI. Section~\hyperref[sec:eval]{Evaluations} details the experimental implementation and discusses the evaluation results. Finally, Section~\hyperref[sec:con]{Conclusion} concludes the paper with a summary of the research findings and potential directions for future research.

\section*{Preliminaries}\phantomsection\label{sec:pre}
A steganographic system can be characterised by several defining properties, including capacity, fidelity, secrecy and robustness. The relative significance of each property depends on the specific application for which the system is intended. We begin by discussing the primary properties typically associated with steganographic systems, and then briefly summarise the scope of this study in relation to each of these properties.

\subsection*{Capacity}
Capacity refers to the number of bits that a steganographic system can embed within a given medium. In general, a message can be mapped into a sequence of symbols drawn from an alphabet $\mathcal{S}$, where each symbol can be represented by $\log_2 \|\mathcal{S}\|$ bits. A zero-bit system determines whether or not a specific symbol is present within a medium. The evaluation of zero-bit systems often involves assessing the false alarm rate, which represents the probability that a symbol will be detected in a medium when, in fact, no symbol is actually present. In contrast, a multi-bit system encodes multiple bits of information within a medium, allowing for more practical applications. The evaluation of multiple-bit systems concerns the frequency with which symbols are incorrectly decoded. This study develops both zero-bit and multi-bit steganographic systems for multimodal media.

\subsection*{Fidelity}
Fidelity refers to the degree of similarity between the cover and stego versions of the medium. A typical fidelity requirement is based on perceptual similarity, guaranteeing that any distortion caused by the steganographic system is either invisible or inaudible to human perception. The steganographic system should not compromise the integrity of the auditory or visual quality to the point of being noticeable to the listener or viewer. In a broader sense, however, the concept of fidelity can be relaxed to refer to semantic similarity, where the alterations made to the medium may be noticeable but do not change the fundamental meaning or purpose conveyed. This is often exemplified in linguistic steganography, where changes in lexicon, syntax, or even language may result in synonymous expressions or paraphrases that, while recognisable to a human observer, retain the core idea and intent of the original text. For human-centric audiovisual media, fidelity can be broadened to encompass biometric similarity, including face identity for visual content and voice identity for auditory content. This study focuses on both semantic and biometric similarities in the context of multimodal steganography.

\subsection*{Secrecy}
Secrecy refers to the inconspicuousness of a stego medium to an adversary. It is the primary concern in applications such as covert communications, where the objective is to hide the very fact that secret communication is taking place. The process of detecting stego media is known as steganalysis, in which secrecy is often modelled as statistical undetectability. The presence of statistical or contextual anomalies can be indicative, and therefore, may be used to infer the likelihood of a covert communication. At an abstract level, the statistics of the cover media can be seen as a probability distribution of all possible cover media. However, it is inherently difficult to model such a probability distribution, and therefore, simplified models of cover distribution that reflect some expected characteristics are often used in practice. This study analyses statistical deviations through both empirical and learning-based models.

\subsection*{Robustness}
Robustness refers to the ability to reliably decode the message even in the presence of unintentional degradations or malicious operations. It is pivotal in applications such as fact-checking and cyber-forensics, as it enables the tracing of a content’s origin and provides a means to verify whether the media has been fabricated or misrepresented. Central to achieving robustness is the principle of orthogonality, which ensures that the embedded message remains unaffected and detectable despite a range of potential manipulations. When the message is embedded in a domain that is either orthogonal or invariant to the potential attacks, it remains immune to their effects. For example, a visual or auditory medium can be projected into the frequency domain, where the message is embedded in parts of the frequency spectrum that are less likely to be compromised by potential manipulations. This study exploits the orthogonality between the linguistic and audiovisual domains, allowing the message to withstand changes in either the auditory or visual modality.

\begin{figure*}[t!]
\centering
\includegraphics[width=0.99\linewidth]{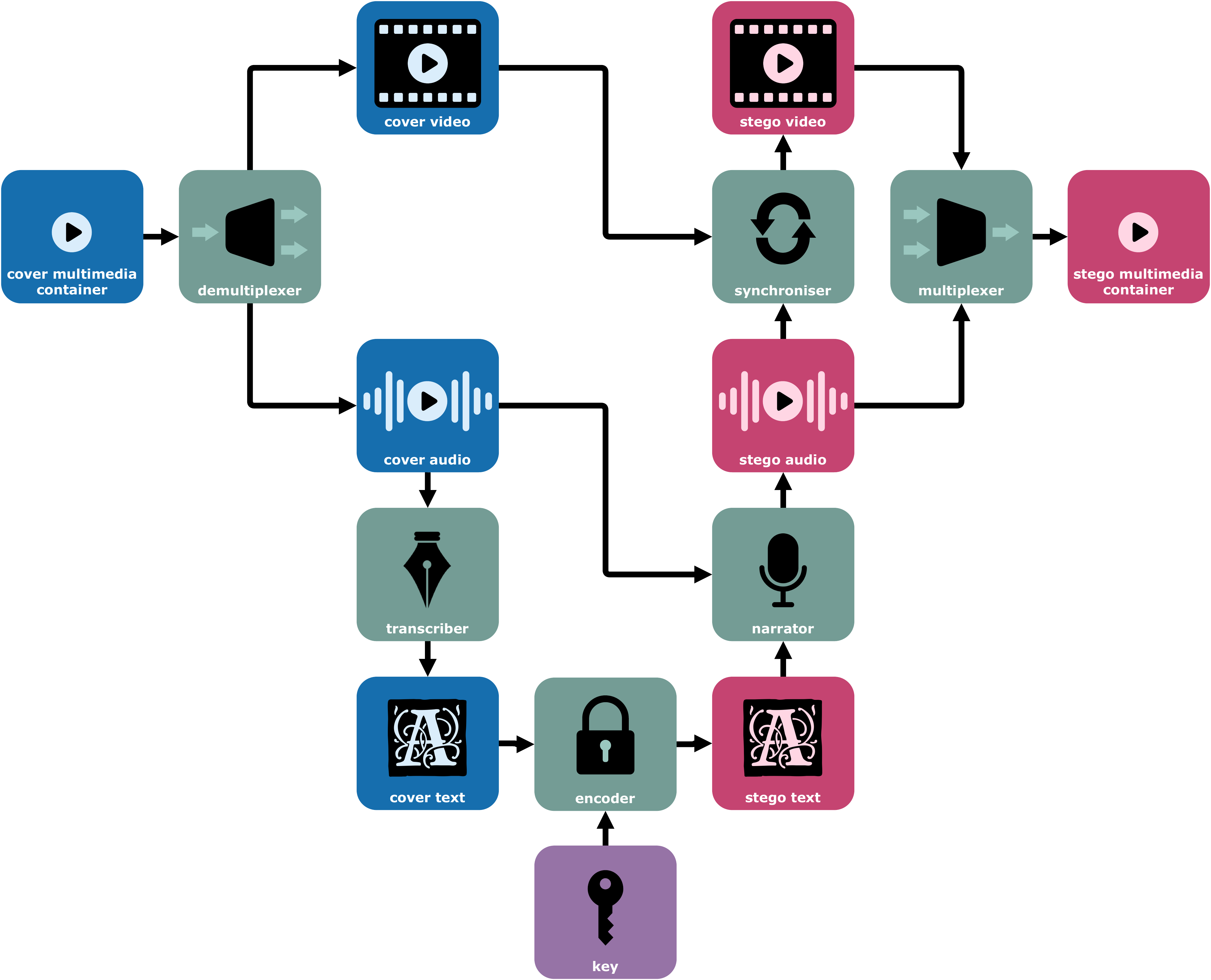}
\caption{Overview of the message encoding process with a shared key, converting a cover multimedia container into a stego multimedia container.}
\label{fig:encoding}
\end{figure*}

\section*{Methodology}\phantomsection\label{sec:method}
In general, a steganographic system consists of an encoding process at the sender's side and a decoding process at the receiver's side. It is assumed that one or more keys are shared between the sender and the receiver through a secure key exchange protocol, with the number of keys depending on the capacity setting. We begin by outlining each process and then describe the key components of the proposed methodology.

\subsection*{Message Encoding Process}
The encoding process begins with demultiplexing the cover multimedia container into cover video and cover audio. Next, the cover audio is transcribed into cover text, which is then encoded into stego text using the shared key. In the zero-bit capacity setting, only one key $K$ is shared. In the multi-bit capacity setting, the key $K$ to be used is selected from a shared key set based on the corresponding message symbol $S$, where the size of the message alphabet $\| \mathcal{S} \|$ matches the size of the key set $\|\mathcal{K}\|$. The stego text is then narrated into stego audio, and the cover video is synchronised with the stego audio. Finally, both stego video and stego audio are multiplexed into a stego multimedia container. An overview of the message encoding process is illustrated in Figure~\ref{fig:encoding}.

\paragraph{Demultiplexing}
   A multimedia container $C$ includes both a video stream $V$ and an audio stream $A$. The container is split into its individual components as follows:
   \begin{equation}
   \{ V, A \} \leftarrow \operatorname{Demux}(C).
   \end{equation}
   This separation allows independent processing of each stream.

\paragraph{Transcription}  
   The cover audio stream $A$ is transcribed into the cover text transcript $T$ using a speech-to-text (STT) model:
   \begin{equation}
   T = \operatorname{STT}(A).
   \end{equation}
   This transcript serves as a linguistic medium for carrying the message.

\paragraph{Encoding}  
   Consider a language generation model prompted to paraphrase the given text. The word sampling process during paraphrasing is parameterised by a shared key $K$, selected based on the intended message symbol. The language generation model paraphrases the cover text transcript $T$ and results in a stego text transcript $T'$:
   \begin{equation}
   T' = \operatorname{Gen}(T; K).
   \end{equation}
   Specifically, the stego key $K$ serves as a pseudo-random seed for selecting a set of keywords (tokens) $\mathcal{W}$ from the dictionary $\mathcal{V}$, with a pre-defined ratio of selection $\delta$ (0.5 by default), where
   \begin{equation}
   \delta = \frac{\| \mathcal{W} \|}{\| \mathcal{V} \|} .
   \end{equation}
   The word sampling process is biased toward generating tokens in the keyword set $\mathcal{W}$. The output of a language generation model is a sequence of logits, representing unnormalised probabilities of selecting each token from the dictionary. Let $z_w$ denote the logit for the token $w$. The probability of selection is biased by adjusting the logit:
  \begin{equation} z'_w = 
  \begin{cases} 
  z_w + \alpha, & \text{if } w \in \mathcal{W}, \\ 
  z_w, & \text{otherwise},
  \end{cases} 
  \end{equation}
  where $\alpha$ is a positive parameter controlling the bias strength (4 by default). The probability for sampling each token is then updated by applying the softmax function:
  \begin{equation} 
  p_w = \frac{\exp(z'_w)}{\sum_{v \in \mathcal{V}} \exp(z'_v)}. 
  \end{equation}
  As long as the size of the keyword set is large enough, it is possible to preserve semantic equivalence between the cover and stego transcripts, with a statistically significant portion of contextual synonyms sampled from the keyword set.

\begin{figure*}[t!]
\centering
\includegraphics[width=0.99\linewidth]{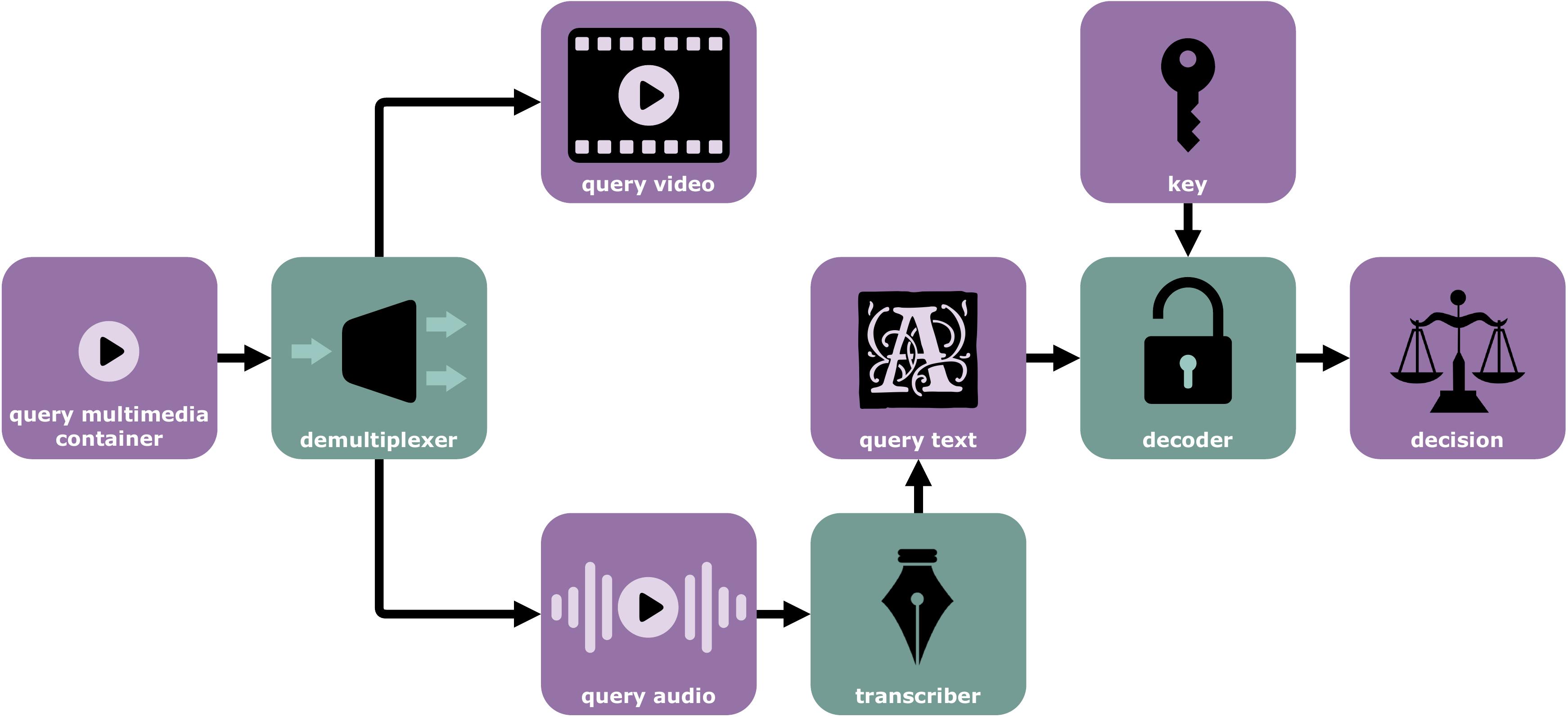}
\caption{Overview of the message decoding process with a shared key, making a binary decision on a query multimedia container.}
\label{fig:decoding}
\end{figure*}

\paragraph{Narration}  
   The stego text transcript $T'$ is converted into the auditory modality using the voice cloned from the cover audio stream $A$ through a text-to-speech (TTS) model:
   \begin{equation}
   A' = \operatorname{TTS}(T', A).
   \end{equation}
   This voice-cloning process allows the resulting stego audio stream to retain the characteristics of the original speaker.

\paragraph{Synchronisation}
	To maintain audiovisual consistency, the stego audio stream $A'$ is synchronised with the cover video stream $V$ using a lip-synchronisation model:
   \begin{equation}
   V' = \operatorname{Sync}(V, A').
   \end{equation}

\paragraph{Multiplexing} 
   Finally, the stego video stream $V'$ and stego audio stream $A'$ are combined into a stego multimedia container $C'$:
   \begin{equation}
   C' = \operatorname{Mux}(V', A').
   \end{equation}
   This stego multimedia container is then transmitted via a channel to the receiver, with the risk of unauthorised manipulation.

\subsection*{Message Decoding Process}
The decoding process begins with demultiplexing the query multimedia container into query video and query audio. Next, the query audio is transcribed into query text, which is then decoded using one or more shared key. In the zero-bit capacity setting, only one key is shared, and the decision is based on applying a threshold to the likelihood that the query text is unbiased, given the number of observed tokens belonging to the keyword set. In the multi-bit capacity setting, however, every shared key is applied, and the decision is made by selecting the key that minimises the likelihood of unbiasedness, given the keyword set seeded by the key. An overview of the message decoding process is illustrated in Figure~\ref{fig:decoding}.

\paragraph{Demultiplexing}
   A query multimedia container $\hat{C}$ consists of a query video stream $\hat{V}$ and a query audio stream $\hat{A}$. The query container is separated into its individual components as follows:
   \begin{equation}
   \{ \hat{V}, \hat{A} \} \leftarrow \operatorname{Demux}(\hat{C}).
   \end{equation}
   
\paragraph{Transcription}  
   The query audio stream $\hat{A}$ is transcribed into a query text transcript $\hat{T}$ using an STT model:
   \begin{equation}
   \hat{T} = \text{STT}(\hat{A}).
   \end{equation}

\begin{figure}[t!]
\centering
\includegraphics[width=0.99\linewidth]{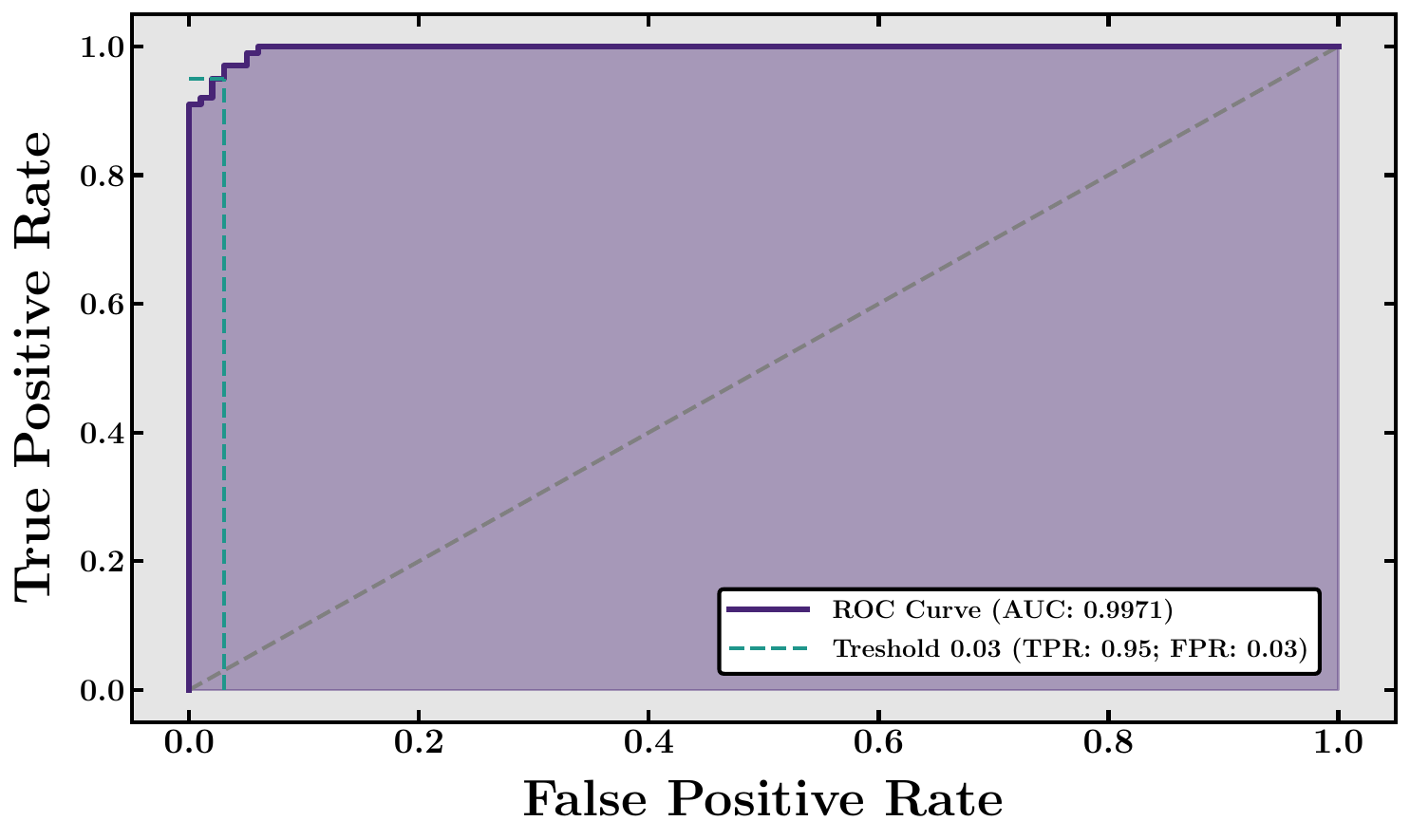}
\caption{Evaluation of accuracy in zero-bit capacity setting.}
\label{fig:accuracy}
\end{figure}

\paragraph{Decoding}
In the zero-bit setting, to verify the presence of the message, the decoding process infers the probability that the query text transcript is unbiased, assessing whether the observed number of keywords in the query text transcript deviates significantly from what would be expected under an unbiased scenario. Let $n$ be the total number of tokens in the query text transcript. The number of tokens belonging to the keyword set $\mathcal{W}$ is computed by:
\begin{equation} 
t = \sum_{w=1}^n \mathbb{I}(z_w \in \mathcal{W}), 
\end{equation}
where $I$ is the indicator function. The probability of observing more than $t$ keyword tokens is given by the survival function of a binomial distribution $\operatorname{Binomial}(n, \delta)$, where $\delta$ is the ratio of keyword selection. Mathematically, the survival function is the complementary cumulative distribution function, as defined by:
   \begin{equation}
   \operatorname{SF}(t) = 1 - \operatorname{CDF}(t) = 1 - \sum_{i=0}^{t} \binom {n}{i} \delta^{i} (1-\delta)^{n-i} .
   \end{equation}
If the probability given by the survival function is lower than a predefined threshold $\theta$ (0.03 by default), the query text transcript is considered to contain the message, as expressed by:
\begin{equation} \text{Decision} = 
\begin{cases} \text{True}, & \text{if } \operatorname{SF}(t) < \theta, \\
 \text{False}, & \text{otherwise}. 
\end{cases} 
\end{equation}
In the multi-bit setting, to determine the intended message symbol, the decoding process is applied over all keys and selects the one that yields the lowest probability, representing the deviation farthest from the expected statistics of unbiased scenario, as expressed by
\begin{equation} 
\text{Decision} = \argmin_{K \in \mathcal{K}} \operatorname{SF}(t_{K}) ,
\end{equation}
where $t_{K}$ is the number of tokens belong to the keyword set seeded by the key $K$. Finally, the key that yields the lowest probability can be mapped back to the message symbol.

\begin{figure}[t!]
\centering
\includegraphics[width=0.99\linewidth]{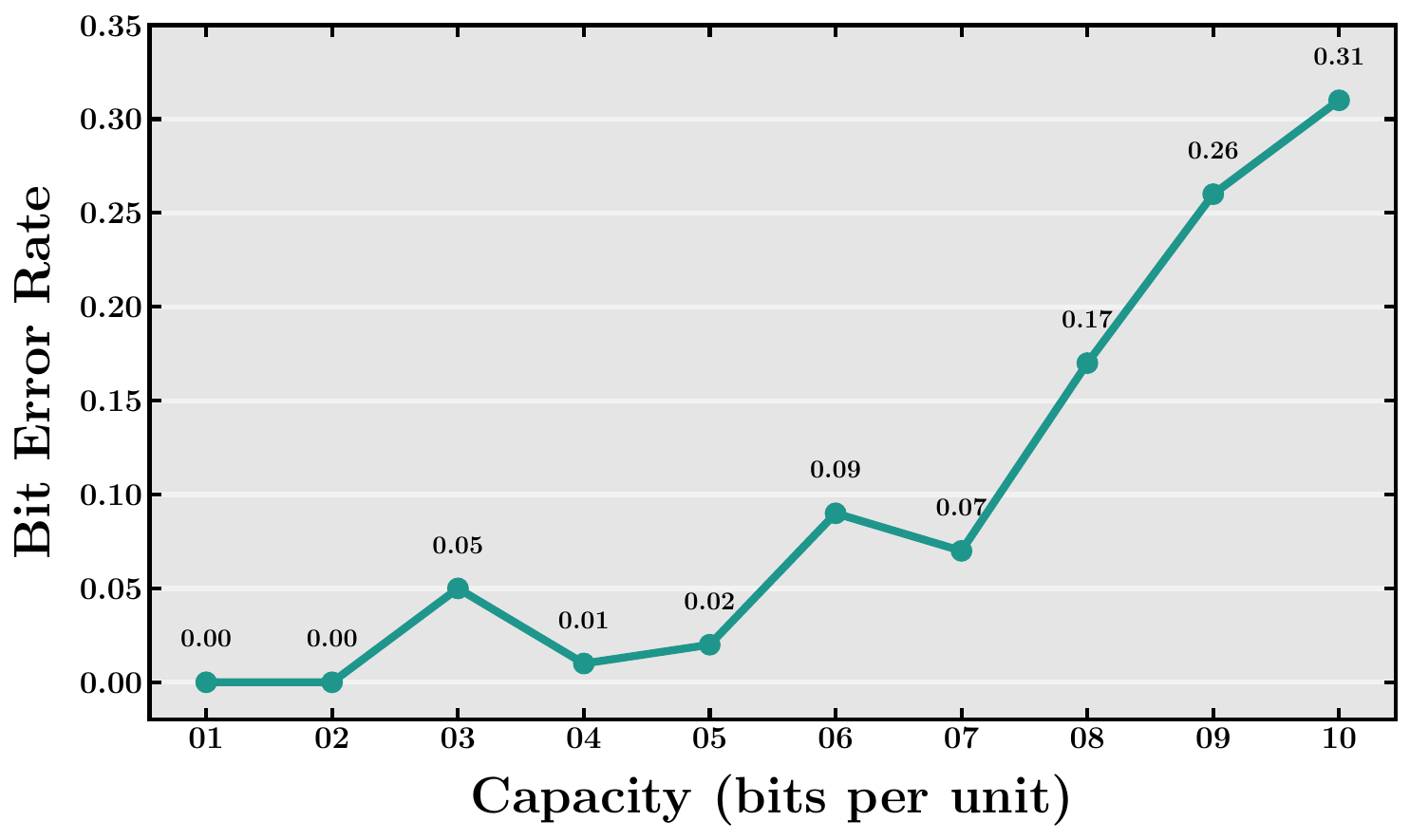}
\caption{Evaluation of accuracy in multi-bit capacity setting.}
\label{fig:capacity}
\end{figure}

\section*{Evaluations}\phantomsection\label{sec:eval}
To validate the concept of the proposed steganographic methodology, we conducted a series of experiments to evaluate various aspects of the system, including capacity, fidelity, secrecy, and robustness.

\subsection*{Experimental Setup} 
The proposed steganographic system involves a sequence of pre-trained machine learning models, each responsible for a specific task in the process. The selected models are all open-source and state-of-the-art within their respective domains, as outlined below:
\begin{itemize}
	\item Transcriber: An open-source multilingual speech-recognition model named Whisper developed by OpenAI~\cite{pmlr-v202-radford23a}.
	\item Narrator: An open-source multilingual text-to-speech model named XTTS developed by Coqui~\cite{casanova24_interspeech}.
	\item Synchroniser: An open-source lip-synchronisation model named Wav2Lip developed by the academic community~\cite{10.1145/3394171.3413532}.
	\item Language Generator: An open-source language generation model named Llama developed by Meta~\cite{grattafiori2024llama3herdmodels}.
\end{itemize}
These models were integrated to enable a fully automated steganographic system capable of generating semantics-aligned identity-preserved audiovisual media. The applied LLama model is version 3.2, with instruction-tuning and 1 billion parameters. At the time of writing, it represents a state-of-the-art release with a minimal model size, making it suitable for deployment and execution on lightweight hardware for reasoning tasks. The temperature of language generation was set to 0.1 to concentrate the sampling process on the most probable tokens, resulting in less variety in responses but higher coherence. The evaluations on zero-bit capacity, fidelity and robustness were conducted on a 9-second speech by C. E. Shannon from the MIT centennial film `The Thinking Machine'~\cite{infiniteMIT}, which served as the cover medium. The evaluations on multi-bit capacity and secrecy were conducted on 100 paragraphs of approximately 50 words each from `Alice’s Adventures in Wonderland' by Lewis Carroll.

\begin{figure}[t!]
\centering
\includegraphics[width=0.99\linewidth]{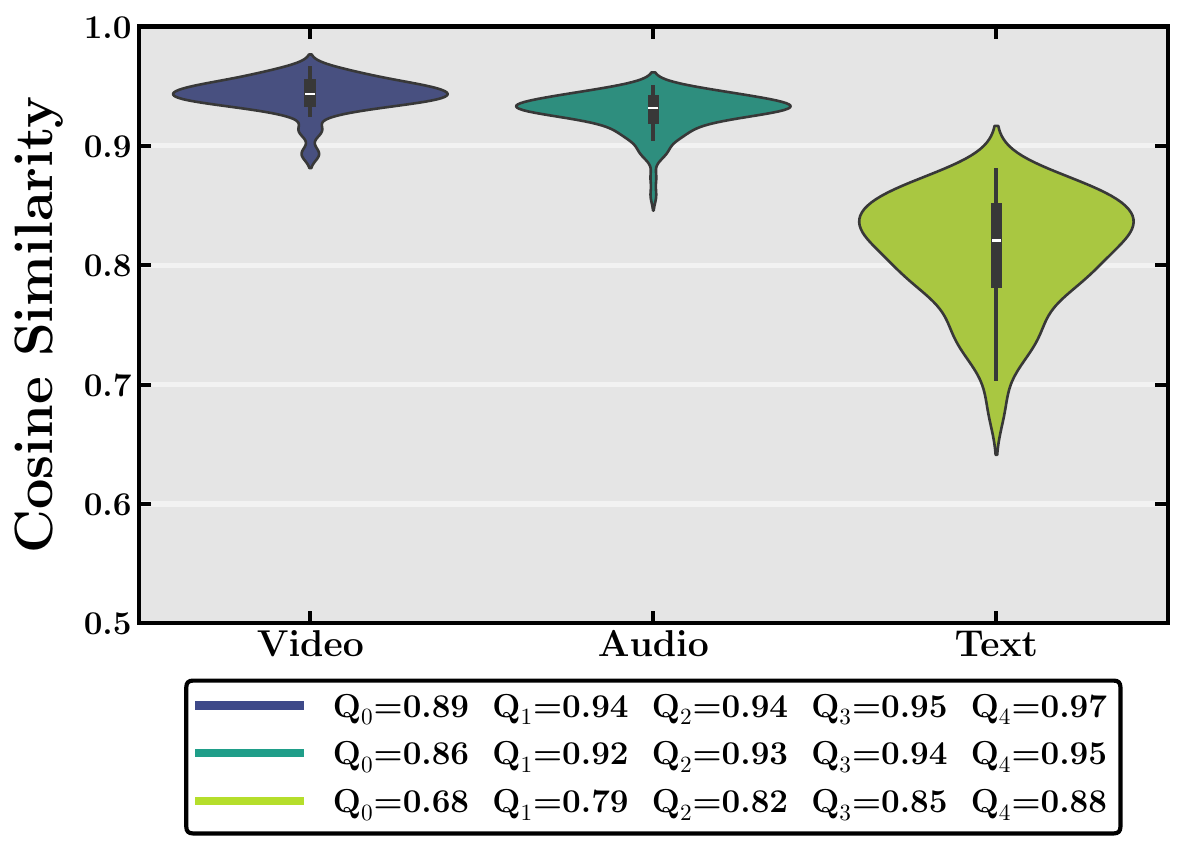}
\caption{Evaluation of fidelity in terms of biometric and semantic similarities.}
\label{fig:fidelity}
\end{figure}

\subsection*{Evaluation of Zero-Bit Capacity}
Figure~\ref{fig:accuracy} shows the accuracy of the steganographic system under a zero-bit setting, where the objective is to determine whether a mark is present or absent in a given query medium. We generated 100 marked transcripts from the cover transcript using varied stego keys and 100 unmarked transcripts by paraphrasing the cover transcript without biasing the word sampling process. The receiver operating characteristic (ROC) curve, with an area under the curve (AUC) value of 0.9971, demonstrates the system's high discriminative power in distinguishing between marked and unmarked media. When the threshold $\theta$ was set to 0.03, the true positive rate (TPR) was 0.95 and the false positive rate (FPR) was 0.03, reflecting the system's ability to reliably detect marked media while maintaining a low probability of raising false alarms.

\subsection*{Evaluation of Multi-Bit Capacity}
Figure~\ref{fig:capacity} presents the relationship between capacity and bit error rate (BER) under a multi-bit setting, where the objective is to determine the most likely message symbol from a given query medium. For each capacity (or bandwidth) setting, we embedded a random message symbol into each cover paragraph using the corresponding stego key and extracted the message symbol from each stego paragraph by comparing the resulting probabilities across all stego keys. The number of stego keys was 2 raised to the power of capacity. For lower capacity settings from 1 to 7 bits per unit, the BER remained relatively low, ranging from 0.00 to 0.09. However, as capacity increased beyond 7 bits per unit, the BER began to rise more significantly, reaching 0.31 at a capacity of 10 bits per unit. This trend suggests a trade-off between the amount of data embedded and the accuracy of data extracted, where a higher bandwidth may lead to a higher probability of collision.

\begin{figure}[t!]
\centering
\includegraphics[width=0.99\linewidth]{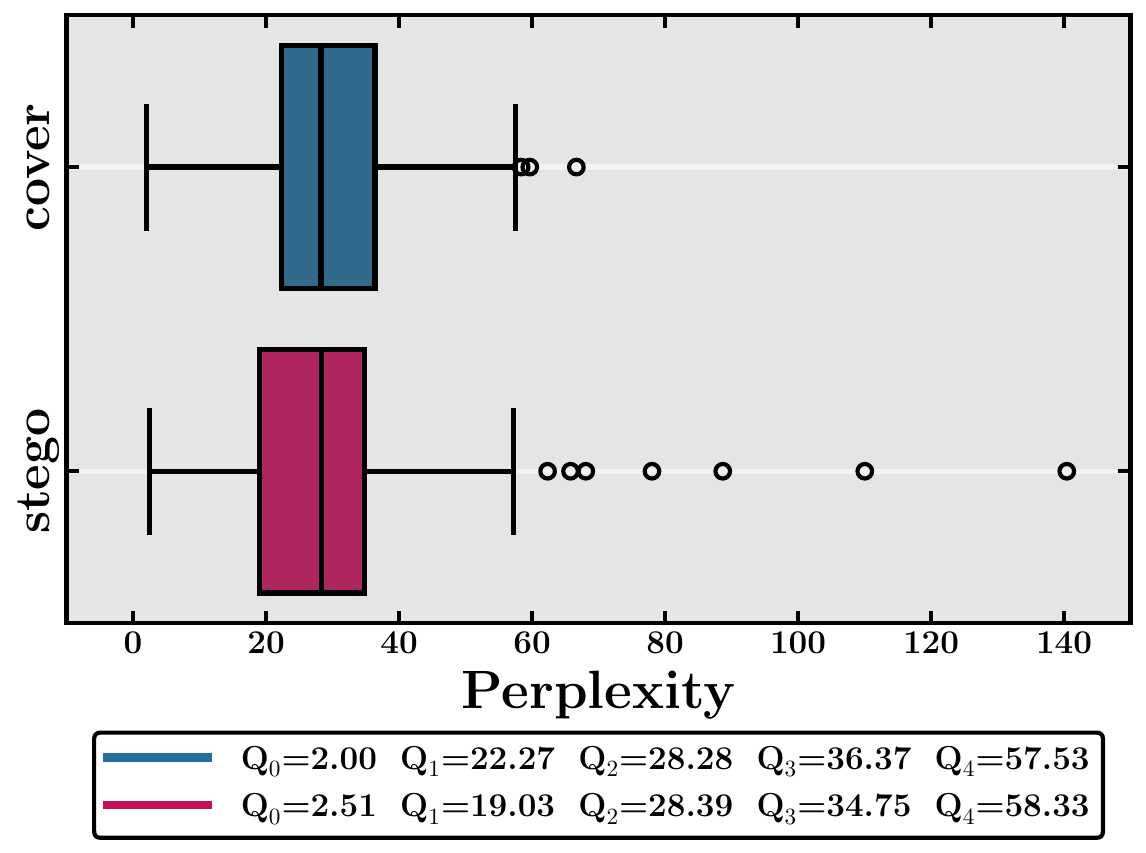}
\caption{Evaluation of secrecy with respect to perplexity.}
\label{fig:perplexity}
\end{figure}

\subsection*{Evaluation of Fidelity}
Figure~\ref{fig:fidelity} presents the cosine similarity across three modalities: video, audio and text, evaluating the alignment of latent embeddings in 100 stego media compared to the cover medium. We converted video streams into face identity embeddings with FaceNet~\cite{7298682}, audio streams into voice identity embeddings with SpeechBrain~\cite{JMLR:v25:24-0991}, and text transcripts into contextual word embeddings with BERT~\cite{devlin-etal-2019-bert}. The quartile $Q_0$ denotes the minimum, $Q_1$ the 25th percentile, $Q_2$ the median, $Q_3$ the 75th percentile and $Q_4$ the maximum. For video streams, the face identity similarities were high with a median of 0.94. For audio streams, the voice identity similarities also remained high with a median of 0.93. For text transcripts, the contextual word similarities were slightly lower and exhibited wider variations with a median of 0.82. This could be improved through the art of prompt engineering, which encourages the language model to more strictly mirror the original text or echo it, rather than loosely paraphrasing, thereby limiting the degree of variation in the generated text.

\subsection*{Evaluation of Secrecy}
\begin{figure}[t!]
\centering
\includegraphics[width=0.99\linewidth]{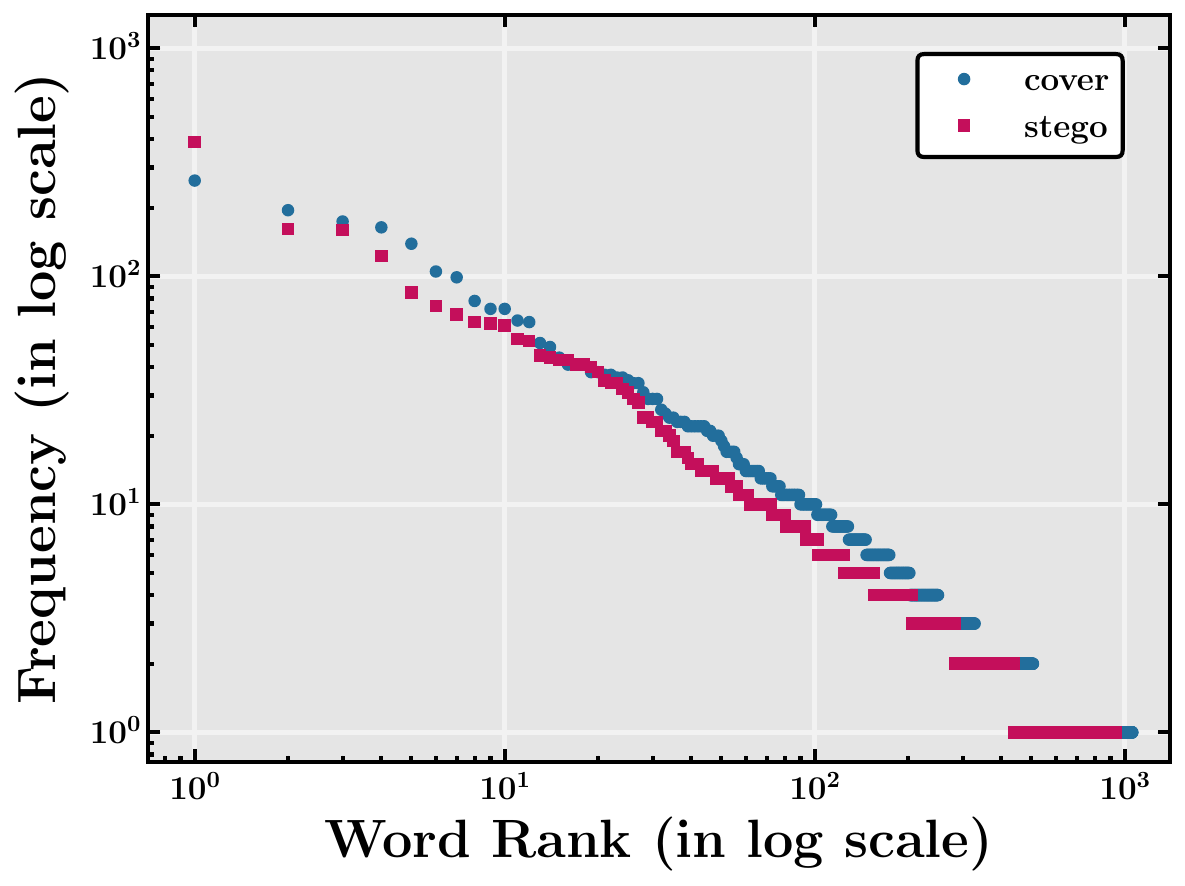}
\caption{Evaluation of secrecy with respect to Zipf's law.}
\label{fig:zipf}
\end{figure}

At first glance, the challenge of steganalysis may appear to overlap with generative AI detection, as the proposed system involves a chain of AI to generate stego objects. However, as generative AI becomes increasingly prevalent in society, AI-synthesised content alone does not necessarily imply concealed communication. Consequently, classifying all AI-generated content as stego objects would be impractical. Due to the lack of existing steganalysis tools tailored to our proposed system, we evaluate the statistical deviations between cover and stego texts. Figure~\ref{fig:zipf} shows the frequency distribution of words in the cover and stego texts, plotted on a log-log scale. The distribution for both cover and stego texts follows a similar pattern, with a steep decline in frequency as rank increases, typical of a Zipfian distribution~\cite{Zipf:1935aa}. Figure~\ref{fig:perplexity} presents the perplexity distribution for both cover and stego texts, measured using the Llama language model. Both cover and stego texts exhibit a similar range and central tendency in perplexity values, although the stego texts contain more outliers. These findings suggest that, as long as the stego texts are not outliers, they can be sampled with the same level of uncertainty as the cover texts, as reflected in their perplexity distribution. The results further support the statistical indistinguishability of the stego content.

\subsection*{Evaluation of Robustness}
\begin{figure}[t!]
\centering
\includegraphics[width=0.99\linewidth]{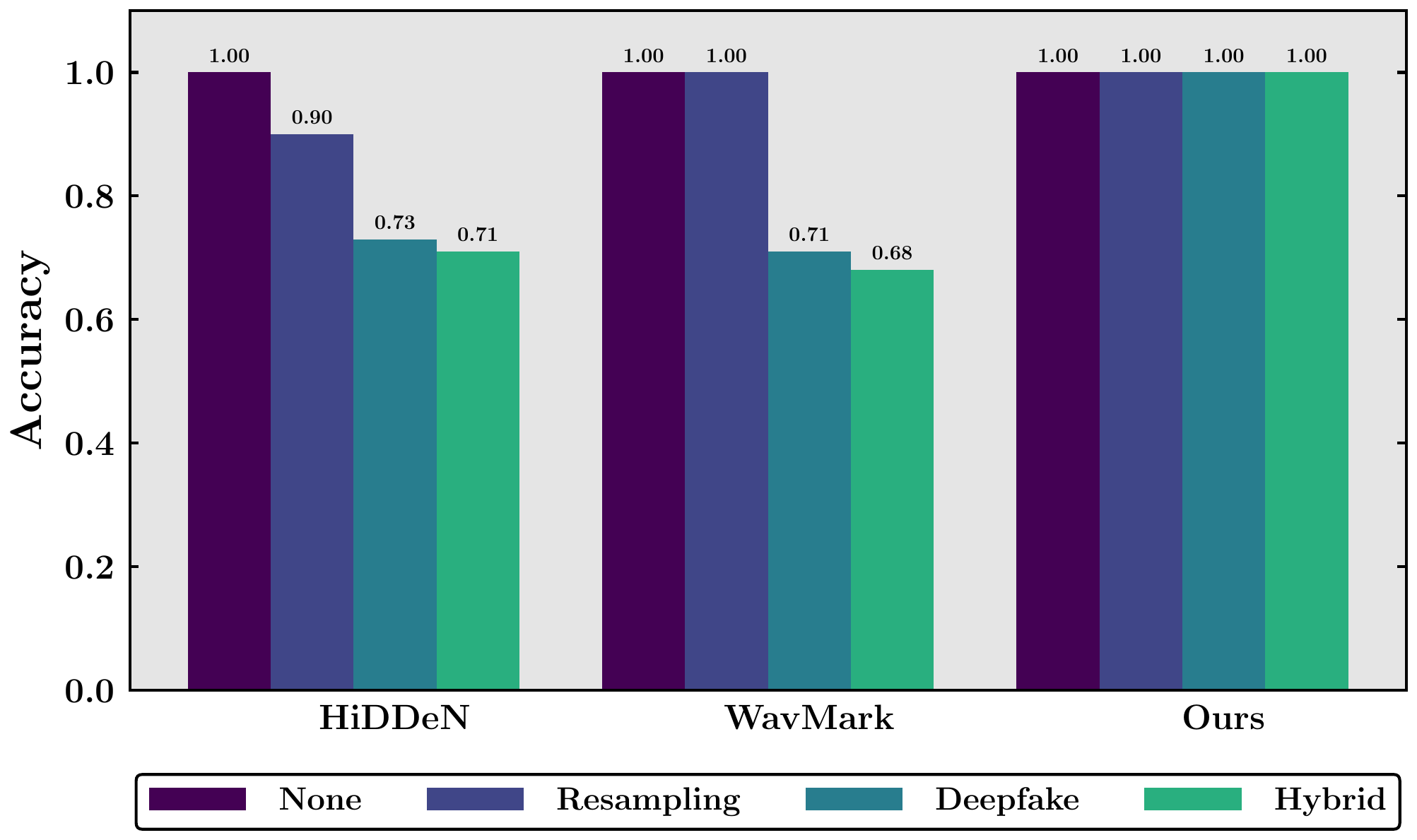}
\caption{Evaluation of robustness across various scenarios.}
\label{fig:robustness}
\end{figure}

\begin{figure*}[t!]
\centering
\includegraphics[width=0.99\linewidth]{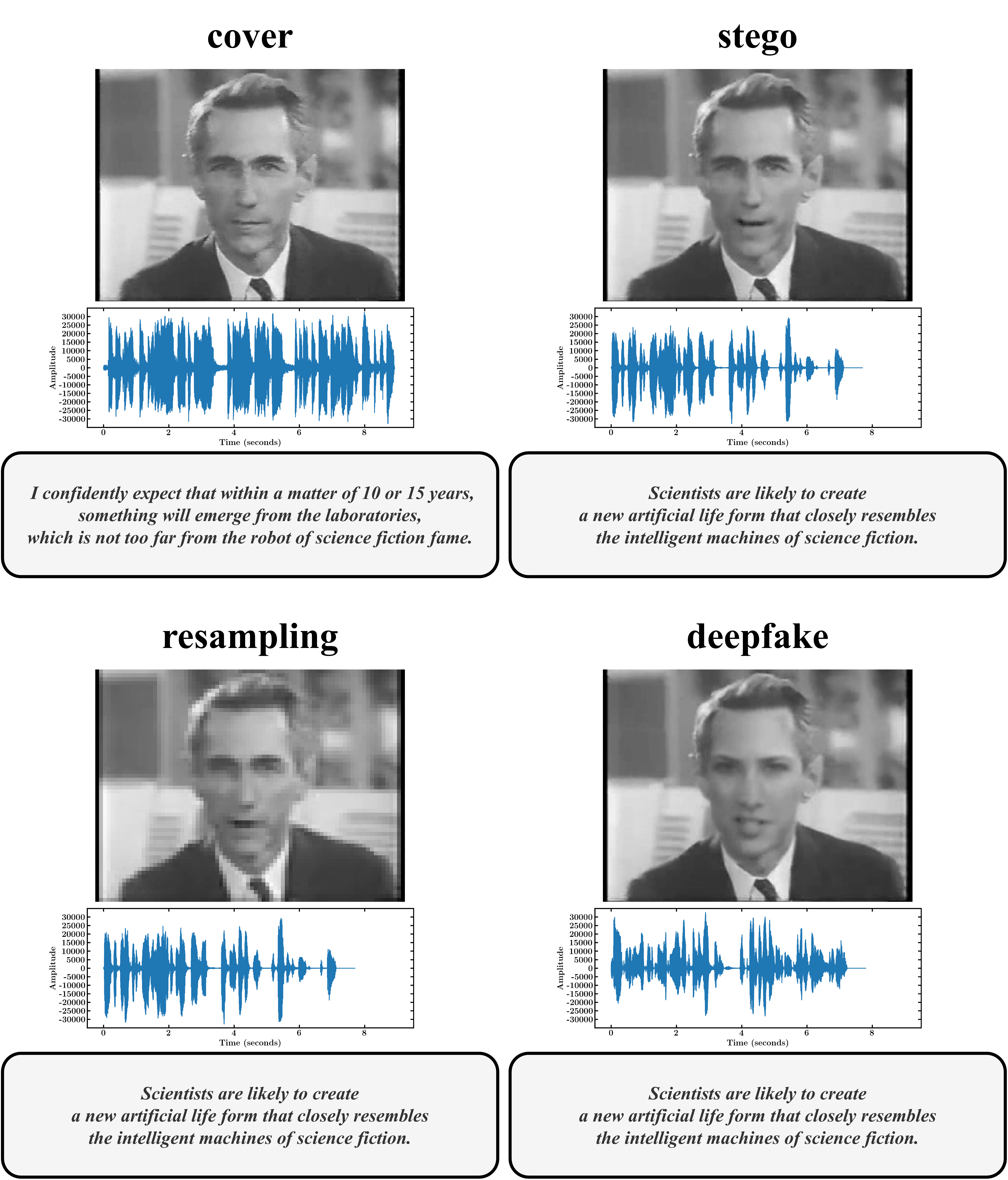}
\caption{Visualisation of cover, stego, resampling and deepfake versions of media across visual, auditory and linguistic modalities.}
\label{fig:vis_example}
\end{figure*}

Figure~\ref{fig:robustness} compares the robustness of the proposed steganographic system with the visual steganographic system HiDDeN~\cite{Zhu:2018aa} and the auditory steganographic system WavMark~\cite{chen2024wavmarkwatermarkingaudiogeneration} in terms of message extraction accuracy (complementary to BER). The evaluation was conducted under four conditions: no manipulation, resampling, deepfake, and hybrid, which combines both resampling and deepfake manipulations. The resampling algorithm reduced the display resolution to 25\%, from 320 × 240 to 80×60 pixels, halved the video frame rate from 14 to 7 frames per second (fps), and halved the audio sample rate from 24 to 12 kilohertz (kHz). Deepfake video streams were generated using the face-swapping model SimSwap~\cite{10.1145/3394171.3413630}, with 100 faces from the FaceForensics++ dataset~\cite{9010912}. Deepfake audio streams were generated using the voice-cloning model XTTS~\cite{casanova24_interspeech}, with 100 voices from the LibriSpeech corpus~\cite{7178964}. Our system was configured to operate at a capacity of one bit per audiovisual content. When no modification was applied, all three systems achieved perfect accuracy, confirming their ability to reliably communicate messages in an ideal setting. However, when subjected to resampling, HiDDeN exhibited a slight decline in accuracy, dropping to 0.90, whereas WavMark and our system remained unaffected. Under deepfake transformations, both HiDDeN and WavMark experienced further performance degradation, with accuracy dropping to 0.73 and 0.71, respectively, while our method maintained perfect robustness. The hybrid scenario, which introduced both resampling and deepfake manipulations, resulted in the lowest accuracy for HiDDeN at 0.71 and WavMark at 0.68, yet our system remained entirely resilient. Overall, these results demonstrate the vulnerability of existing systems to various forms of distortion, whereas our method consistently demonstrates superior robustness across all tested scenarios. Figure~\ref{fig:vis_example} visualises a pair of cover and stego media, along with the resampling and deepfake variants. It can be observed that while the cover transcript was altered by the steganographic system, the core semantics were preserved in the stego transcript. For instance, the text changed from \textit{`I confidently expect that within a matter of 10 or 15 years, something will emerge from the laboratories, which is not too far from the robot of science fiction fame'} to \textit{`Scientists are likely to create a new artificial life form that closely resembles the intelligent machines of science fiction'}.


\section*{Conclusion}\phantomsection\label{sec:con}
In conclusion, this study investigates the vulnerability of steganography to the evolving capabilities of generative AI, with a focus on the risk of synthetic content overwriting hidden messages. Instead of relying on the spatial or temporal domains, messages are embedded in the linguistic domain of audiovisual content based on a chain of multimodal AI. At the core of this chain is a language generation model tasked with paraphrasing the given transcript, while the word sampling process is biased towards a shared set of tokens. The paraphrased transcript is then converted into sound using a voice-cloning model, and the video is aligned with the audio using a lip-synchronisation model. Several aspects of the proposed steganographic system are evaluated, including accuracy, fidelity, secrecy and robustness, across various metrics and conditions. 

The proposed system is inherently composed of a tightly coupled chain of AI components, each dependent on the output of the previous stage. While this sequential collaboration enables coherent synthesis of audiovisual stego media, it also introduces a high degree of interdependence, which may affect system flexibility and fault tolerance. A failure or degradation in an antecedent module can propagate downstream, potentially compromising stability in both encoding and decoding. This architectural choice prioritises semantic alignment but may not represent the optimal strategy in all contexts. An alternative design could involve parallel multimodal embeddings to enhance robustness and enable message extraction and verification across different modalities. In other words, future research may explore the integration of steganographic methods for individual modalities into a unified framework, where they complement each other through the cooperation and interaction of multimodal AI systems.

\IEEEtriggeratref{66}
\bibliography{Transactions-Bibliography/bstcontrol, Bib/bib_spacetime}
\bibliographystyle{Transactions-Bibliography/IEEEtran}

\end{document}